\documentclass[
    ,final            
    ,sort&compress    
    ,numberedheadings 
  ]
  {aipproc}

\usepackage{hyperref}           
\usepackage[latin1]{inputenc}   
\usepackage{latexsym}
\usepackage{amsmath}   
\usepackage{amsfonts}
\usepackage{amssymb}
\usepackage{graphicx}
\usepackage{amsthm}

\layoutstyle{8x11single}

\providecommand{\email}[1]{E-mail:
\href{mailto:#1}{\textup{\texttt{#1}}}}

 \providecommand{\dprod}{\! \cdot \!}%
 \providecommand{\wprod}{\! \wedge \!}
 
\begin{document}

\title{Geometric Drive of the Universe's Expansion}

 \classification{98.80.Jk, 04.50.+h, 02.40.Dr}
  \keywords      {Hubble expansion, Euclidean relativity, dark matter}

\author{José B. Almeida}{
  address={Universidade do Minho, Physics Department, Braga,
Portugal. \email{bda@fisica.uminho.pt} }}

\begin{abstract}
What if physics is just the way we perceive geometry? That is, what
if geometry and physics will one day become one and the same
discipline? I believe that will mean we will at last really
understand physics, without postulates other than those defining the
particular space where the physics play is performed. In this paper
I use 5-dimensional spacetime as a point of departure and make a
very peculiar assignment between coordinates and physical distances
and time. I assume there is an hyperspherical symmetry which is made
apparent by assigning the hypersphere radius to proper time and
distances on the hypersphere to usual 3-dimensional distances. Time,
or Compton time to distinguish from cosmic time is the 0th
coordinate and I am able to project everything into 4-dimensions by
imposing a null displacement condition.

Surprisingly nothing else is needed to explain Hubble's expansion
law without any appeal to dark matter; an empty Universe will expand
naturally at a flat rate in this way. I then discuss the
perturbative effects of a small mass density in the expansion rate
in a qualitative way; quantitative results call for the solution of
equations that sometimes have not even been clearly formulated and
so are deferred to later work. A brief outlook of the consequences
an hyperspherical symmetry has for galaxy dynamics allows the
derivation of constant rotation velocity curves, again without
appealing to dark matter.

An appendix explains how electromagnetism is made consistent with
this geometric approach and justifies the fact that photons must
travel on hypersphere circles, to be normal to proper time.

\end{abstract}

\maketitle

\section{Introduction}

The validity of any theory and its usefulness stem from the
correctness of the predictions it allows; this is an unquestionable
truth for all physicists and for the public in general. The elegance
of a theory, however, is usually associated to a small number of
principles or postulates and to a small set of mathematical
equations, even if these turn out mathematically intricate and
difficult to solve. This has been the case with General Relativity
(GR) for many years, a theory which many physicists see as the
paradigm of elegance. In spite of the unescapable validity of GR in
celestial mechanics and laboratory experiments, the situation is not
as clear in cosmology. The frustration of all known attempts to
unify GR with Quantum Mechanics and the Standard Model of particle
physics is another motivation for many serious people to burn their
eyelashes in the search for some alternative way of formulating a
new all encompassing theory.

In this work we will discuss geometry under the assumption that a
well chosen geometry will allow, one day, the derivation of all the
equations of physics from purely geometrical relations. This is, to
a great extent, a question of the author's personal faith without
too much evidence to support it at the present time, but enough to
motivate his continued search. If the assumption that physics is
born out of geometry is true, then what we have to do is start off
with the appropriate space, make the correct assignments between
coordinates and physical entities and formulate the equations
resulting from space symmetries and other space properties; these
equations shall be the same as we encounter in physics. In previous
work \cite{Almeida04:4} it was shown that hyperbolic 5-dimensional
space, also known as 5-dimensional spacetime, can generate
4-dimensional space without a metric by the condition of null
displacement. This 4D space acquires a metric by promoting one of
the coordinates to interval; depending on the choice of coordinate
one can obtain either the usual GR space or an Euclidean 4D space
designated as 4-Dimensional Optics (4DO) in view of the similarities
with standard 3-dimensional optics. Mapping of geodesics between the
two spaces can be done for all static metrics, as we will show
below; it is not clear at present if the same operation is possible
in some cases for non-static metrics, although it seems very likely
that it is not. However, many interesting cases in GR are governed
by a static metric and we can easily analyse these in 4DO to gain a
different perspective. Einstein's equations cannot be applied in 4DO
and a suitable replacement was proposed in the cited paper, which
leads to similar results in many cases but not in extreme ones.

The purpose of this paper is to show how 4DO can be used to explain
a flat rate expansion of the Universe under zero mass density. When
one of the coordinates of 4DO is associated with the radius of an
hypersphere this coordinate takes the physical meaning of proper
time and flat rate expansion becomes a direct consequence of
geometry. The basic principles involved have been explained in
another paper \cite{Almeida04:1} but the formulation is now cleaner
than the original one. The usual 3 spatial coordinates are then
associated with arc lengths on the hypersphere surface. The metric
of Euclidean 4-space in hyperspherical coordinates is dependent on
the hypersphere radius (proper time) which precludes its direct
mapping into a GR metric; mapping would be possible by resorting to
Cartesian coordinates at the expense of a difficult interpretation
of their significance. We will also discuss the influence of
non-zero mass density to show that small curvature and cosmological
constants are expected. This conclusion can be reached independently
of the set of equations used to find the metric of space with
uniform mass density. Schwarzschild's metric is PPN equivalent to
the exponential metric proposed in both cited papers and
consequently it is irrelevant which one is chosen if only first
order approximation is envisaged.

Dark matter has been postulated not only to explain the rate of
expansion in the Universe but also to account for the incredible
orbital velocities found in spiral galaxies. This is a subject which
cannot be properly addressed in this short presentation; galaxy
dynamics is a difficult subject which the author did not investigate
properly but, also in this case, the postulate of 4DO in connection
with an hyperspherical Universe seems to provide a qualitative
explanation for the observations. We will give a brief indication of
what may become an interesting subject for further work.

\section{Dynamics in 5D spacetime}

In this section we characterize 5-dimensional spacetime and
introduce the pertinent geometric algebra, ${G}_{4,1}$. For a
comprehensive introduction to geometric algebra readers are referred
to the two excellent books \cite{Doran03,Hestenes84}; here we will
assume some familiarity with this tool.

This paper is about geometry and its relation to physics, which
poses a problem with units right from the start. Geometry only cares
about distances and angles, while physics uses a plethora of
different units. Any parallel between the two fields must solve the
units question right from the start. We note that, at least for the
macroscopic world, physical units can all be reduced to four
fundamental ones; we can, for instance, choose length, time, mass
and electric charge as fundamental, as we could just as well have
chosen others. Measurements are then made by comparison with
standards; of course we need four standards, one for each
fundamental unit. But now note that there are four fundamental
constants: Planck constant $(\hbar)$, gravitational constant $(G)$,
speed of light in vacuum $(c)$ and proton electric charge $(e)$,
with which we can build four standards for the fundamental units.
\begin{table}[bt]
\caption{\label{t:factors}Standards for non-dimensional units used
in the text; $\hbar \rightarrow$ Planck constant divided by $2 \pi$,
$G \rightarrow$ gravitational constant, $c \rightarrow$ speed of
light and $e \rightarrow$ proton charge.}
\begin{tabular}{c c c c}
\hline \tablehead{1}{c}{b}{Length} & \tablehead{1}{c}{b}{Time}
& \tablehead{1}{c}{b}{Mass} & \tablehead{1}{c}{b}{Charge} \\
\hline 
$\displaystyle \sqrt{\frac{G \hbar}{c^3}} $ & $\displaystyle
\sqrt{\frac{G \hbar}{c^5}} $  & $\displaystyle \sqrt{\frac{ \hbar c
}{G}} $  & $e$\\ \hline
\end{tabular}
\end{table}
Table \ref{t:factors} lists the standards of this units' system,
frequently called Planck units, which the author prefers to
designate by non-dimensional units. In this system all the
fundamental constants, $\hbar$, $G$, $c$, $e$, become unity, a
particle's Compton frequency, defined by $\nu = mc^2/\hbar$, becomes
equal to the particle's mass and the frequent term ${GM}/({c^2 r})$
is simplified to ${M}/{r}$. We can, in fact, take all measures to be
non-dimensional, since the standards are defined with recourse to
universal constants; this will be our posture. Geometry and physics
become relations between pure numbers, vectors, bivectors, etc., but
the geometric concept of distance is needed only for graphical
representation.

Another problem we have to tackle is one of notation. Since we work
in 5 dimensions but need also to consider 4-dimensional and
3-dimensional subspaces, we introduce an indexing convention which
allows us to recognize immediately to which space or subspace each
index refers. The following diagram shows the index naming
convention used in this paper.

\vspace{12pt} \centerline{\includegraphics[scale=0.8]{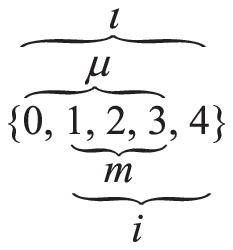}}
\vspace{12pt}

\noindent Indices in the range $\{0,4\}$ will be denoted with Greek
letters $\iota, \kappa, \lambda.$ Indices in the range $\{0,3\}$
will also receive Greek letters but chosen from $\mu, \nu, \xi.$ For
indices in the range $\{1,4\}$ we will use Latin letters $i, j, k$
and finally for indices in the range $\{1,3\}$ we will use also
Latin letters chosen from $m, n, o.$ Einstein's summation convention
will be adopted, as well as the compact {notation} for partial
derivatives $\partial_\iota =
\partial/\partial x^\iota.$ When convenient we will also make the
assignments justified in \citet{Almeida04:4}: $x^0 \equiv t$ and
$x^4 \equiv \tau.$ The squares of coordinates will be denoted by
enclosing in parenthesis, to avoid confusion with superscript
indices, but the same procedure will not be needed for $t, \tau$ and
spherical coordinates.

The {geometric algebra} $G_{4,1}$\index{G(4,1)@$G_{4,1}$} of the
hyperbolic {5-dimensional} space we want to consider is generated by
the {frame} of {orthonormal} vectors $\sigma_\iota $ verifying the
relations
\begin{align}
    & (\sigma_0)^2  = -1,\\ & \sigma_0 \sigma_i + \sigma_i
    \sigma_0
    =0,\\
    & \sigma_i
    \sigma_j + \sigma_j \sigma_i  = 2 \delta_{i j}.
\end{align}
We will simplify the {notation} for {basis} vector products using
multiple {indices}, i.e.\ $\sigma_\iota \sigma_\kappa \equiv
\sigma_{\iota\kappa}.$ The algebra is 32-dimensional and is spanned
by the {basis}
\begin{itemize}
\item 1 scalar, { $1$},
\item 5 vectors, { $\sigma_\iota$},
\item 10 bivectors (area), { $\sigma_{\iota\kappa}$},
\item 10 trivectors (volume), { $\sigma_{\iota\kappa\lambda}$},
\item 5 tetravectors (4-volume), { $\mathrm{i} \sigma_\iota $},
\item 1 {pseudoscalar} (5-volume), { $\mathrm{i} \equiv
\sigma_{01234}$}.
\end{itemize}
Several elements of this {basis} square to unity:
\begin{equation}
    (\sigma_i)^2 =  (\sigma_{0i})^2=
    (\sigma_{0i j})^2 =(\mathrm{i}\sigma_0)^2 =1;
\end{equation}
and the remaining square to $-1:$
\begin{equation}
    \label{eq:negative}
    (\sigma_0)^2 = (\sigma_{ij})^2 = (\sigma_{ijk})^2 =
    (\mathrm{i}\sigma_i)^2 = \mathrm{i}^2=-1.
\end{equation}
Note that the {pseudoscalar} $\mathrm{i}$ commutes with all the
other {basis} elements, while being a square root of $-1$, and plays
the role of the scalar imaginary in complex algebra.

The geometric product of any two vectors $a = a^\iota \sigma_\iota$
and $b = b^\kappa \sigma_\kappa$ is evaluated making use of the
distributive property
\begin{equation}
    ab = \left(-a^0 b^0 + \sum_i a^i b^i \right) + \sum_{\iota \neq \kappa}
    a^\iota b^\kappa \sigma_{\iota \kappa};
\end{equation}
and we notice it can be decomposed into a symmetric part, a scalar
called the inner or interior product, and an anti-symmetric part, a
bivector called the outer or exterior product.
\begin{equation}
    ab = a \dprod b + a \wprod b,~~~~ ba = a \dprod b - a \wprod b.
\end{equation}
Reversing the definition one can write inner and outer products as
\begin{equation}
    a \dprod b = \frac{1}{2}\, (ab + ba),~~~~ a \wprod b = \frac{1}{2}\, (ab -
    ba).
\end{equation}
When a vector is operated with a multivector the inner product
reduces the grade of each element by one unit and the outer product
increases the grade by one. By convention the inner product of a
vector and a scalar produces a vector.

We will encounter exponentials with multivector exponents; two
particular cases of exponentiation are specially important. If $u$
is such that $u^2 = -1$ and $\theta$ is a scalar
\begin{equation}
\begin{split}
   \mathrm{e}^{u \theta} =& 1 + u \theta -\frac{\theta^2}{2!} - u
    \frac{\theta^3}{3!} + \frac{\theta^4}{4!} + \ldots  \\
    =& 1 - \frac{\theta^2}{2!} +\frac{\theta^4}{4!}- \ldots \{=
    \cos \theta \} \\
    &+ u \theta - u \frac{\theta^3}{3!} + \ldots \{= u \sin
    \theta\}\nonumber \\
    =&  \cos \theta + u \sin \theta.
\end{split}
\end{equation}
Conversely if $h$ is such that $h^2 =1$
\begin{equation}
\begin{split}
    \mathrm{e}^{h \theta} =& 1 + h \theta +\frac{\theta^2}{2!} + h
    \frac{\theta^3}{3!} + \frac{\theta^4}{4!} + \ldots  \\
    =& 1 + \frac{\theta^2}{2!} +\frac{\theta^4}{4!}+ \ldots \{=
    \cosh \theta \}\\
    &+ h \theta + h \frac{\theta^3}{3!} + \ldots \{= h \sinh
    \theta\}  \\
    =&  \cosh \theta + h \sinh \theta.
\end{split}
\end{equation}
The exponential of bivectors is useful for defining rotations; a
rotation of vector $a$ by angle $\theta$ on the $\sigma_{12}$ plane
is performed by
\begin{equation}
    a' = \mathrm{e}^{\sigma_{21} \theta/2} a
    \mathrm{e}^{\sigma_{12} \theta/2}= \tilde{R} a R;
\end{equation}
the tilde denotes reversion and reverses the order of all products.
As a check we make $a = \sigma_1$
\begin{eqnarray}
    \mathrm{e}^{-\sigma_{12} \theta/2} \sigma_1
    \mathrm{e}^{\sigma_{12} \theta/2} &=&
    \left(\cos \frac{\theta}{2} - \sigma_{12}
    \sin \frac{\theta}{2}\right) \sigma_1
    \left(\cos \frac{\theta}{2} + \sigma_{12} \sin
    \frac{\theta}{2}\right)\nonumber \\
    &=& \cos \theta \sigma_1 + \sin \theta \sigma_2.
\end{eqnarray}
Similarly, if we had made $a = \sigma_2,$ the result would have been
$-\sin \theta \sigma_1 + \cos \theta \sigma_2.$

If we use $B$ to represent a bivector belonging to {Euclidean}
4-space and define its norm by $|B| = (B \tilde{B})^{1/2},$ a
general rotation is represented by the {rotor}
\begin{equation}
    R \equiv \mathrm{e}^{-B/2} = \cos\left(\frac{|B|}{2}\right) -  \frac{B}{|B|}
    \sin\left(\frac{|B|}{2}\right).
\end{equation}
The rotation angle is $|B|$ and the rotation plane is defined by
$B.$ A {rotor} is defined as a {unitary} even multivector (a
multivector with even grade components only) which squares to unity;
we are particularly interested in rotors with bivector components.
It is more general to define a rotation by a plane (bivector) then
by an axis (vector) because the latter only works in 3D while the
former is applicable in any dimension.

The space spanned by frame vectors $\sigma_\iota$ is flat; its
geodesics are straight lines and we can define an elementary
displacement on a geodesic by the vector
\begin{equation}
\label{eq:displacement}
    \mathrm{d}x = \sigma_\iota \mathrm{d}x^\iota = \sigma_0 \mathrm{d}x^0
    + \sigma_i \mathrm{d}x^i.
\end{equation}
Collapsing 5-dimensional space into 4 dimensions can be achieved by
a projection; we choose to make this transition by imposing a null
displacement condition, that is, the norm of the displacement vector
must be null;
\begin{equation}
    (\mathrm{d}x)^2 = \mathrm{d}x \dprod
    \mathrm{d}x = 0.
\end{equation}
Introducing \eqref{eq:displacement} above we verify immediately that
\begin{equation}
    (\mathrm{d}x^0)^2 - \sum (\mathrm{d}x^i)^2 = 0;
\end{equation}
and this is equivalent to either of the relations
\begin{align}
    \label{eq:twospaces1}
    (\mathrm{d}x^0)^2 =& \sum (\mathrm{d}x^i)^2;\\
    \label{eq:twospaces2}
    (\mathrm{d}x^4)^2 =& (\mathrm{d}x^0)^2 - \sum (\mathrm{d}x^m)^2.
\end{align}
The former of these relations defines an Euclidean 4-space where
$(\mathrm{d}x^0)^2$ is taken as interval and the latter defines
Minkowski spacetime with $(\mathrm{d}x^4)^2$ as interval. We see by
this construction that Euclidean and Minkowski 4-spaces can be taken
as belonging to the null subspace of 5D spacetime; we can obtain one
or the other, depending on the coordinate that we choose for
interval. Remember though we are only considering displacements
along geodesics, i.e.\ straight lines. A very different approach to
the same subject was used in \cite{Almeida02:2} and the first author
to notice this equivalence was probably Montanus
\cite{Montanus91,Montanus01}.

Everything that was said above is true in geometry and has no
implications for physics until we decide to assign some of the
coordinates to physical entities. Some of those assignments are
carried over from previous work; for instance we have already
established that coordinate $x^0$ is to be taken as {time} and
coordinate $x^4$ as {proper time} \cite{Almeida04:4}; accordingly we
will frequently represent $x^0$ with the letter $t$ and $x^4$ with
the letter $\tau$. We will also simplify the notation for time and
proper time derivatives by writing { $\mathrm{d}f /\mathrm{d}t
\equiv \dot{f}$}; { $\mathrm{d} f/\mathrm{d}\tau \equiv \check{f}$}.

Dividing both members of \eqref{eq:displacement} by $\mathrm{d}t$
one defines a 4-dimensional velocity vector $v$;
\begin{equation}
    \label{eq:euclvelocity}
    \dot{x} = \sigma_0 + \sigma_i \dot{x}^i = \sigma_0 + v.
\end{equation}
If we are in the null displacement subspace $(\dot{x})^2$ is
necessarily null and we recognize that $v$ is unitary
\begin{equation}
    \label{eq:vsquare}
    v \dprod v = \sum (\dot{x}^i)^2 = 1.
\end{equation}
The velocity vector can then be obtained by rotation of any unitary
vector and it is particularly interesting to note that it can be
expressed as a rotation of the $\sigma_4$ frame vector.
\begin{equation}
    v = \tilde{R}\sigma_4 R.
\end{equation}
The rotation angle is a measure of the 3-dimensional, physical,
velocity. A null angle corresponds to a $v$ directed along
$\sigma_4$ and null physical velocity, while a $\pi/2$ angle
corresponds to the maximum physical velocity. The idea that physical
velocity can be seen as the 3D component of a unitary 4D vector has
been explored in several papers but see \cite{Almeida01:4}.

Instead of dividing \eqref{eq:displacement} by $\mathrm{d}t$ we can
divide by $\mathrm{d}\tau$, obtaining
\begin{equation}
    \check{x} =  \sigma_0\check{x}^0 + \sigma_m \check{x}^m +
    \sigma_4;
\end{equation}
squaring the second member and noting that it must be null we obtain
\begin{equation}
(\check{x}^0)^2 - \sum (\check{x}^m)^2 = 1.
\end{equation}
We then define a bivector, called relativistic 4-velocity, by
\begin{equation}
    \upsilon = \sigma_{04}\check{x}^0 + \sigma_{m4} \check{x}^m,
\end{equation}
such that $\upsilon^2 = \upsilon \upsilon =1$. The relativistic
4-velocity is a bivector in this space and not a vector as in
special relativity but it represents the same physical concept; in
particular we note that any 4-velocity can be obtained by a Lorentz
transformation of bivector $\sigma_{04}$.
\begin{equation}
    \upsilon = \tilde{T} \sigma_{04} T,
\end{equation}
where $T$ is of the form $T = \exp(B)$ and $B$ is a bivector whose
plane is normal to $\sigma_4$. Note that $T$ is a pure rotation when
the bivector plane is normal to both $\sigma_0$ and $\sigma_4$.

In order to study dynamics we must introduce bent space by allowing
for non-orthonormed frame vectors;
\begin{equation}
    g_\iota = {n^\kappa}_\iota \sigma_\kappa,
\end{equation}
$g_\iota$ is called the \emph{refractive index frame} and
${n^\kappa}_\iota$ the \emph{refractive index tensor} or simply the
\emph{refractive index.} The designation is borrowed from 3D optics
and the refractive index tensor can be seen as the 5D generalization
of a dielectric refractive index. The definition of frame vectors
with recourse to the orthonormed frame can only be applied to bent
spaces and not to general curved ones but we believe this is
sufficient for expressing all dynamics; most of the derivations that
follow, however, would apply equally well to spaces of general
curvature. We introduce now the reciprocal frame $g^\iota$ such that
\cite{Doran03}
\begin{equation}
    g^\iota \dprod g_\kappa = {\delta^\iota}_\kappa.
\end{equation}

In non-orthonormed frames we define the elementary \emph{optical
displacement} vector
\begin{equation}
\label{eq:opticaldisp}
 \mathrm{d}s = g_\iota \mathrm{d}x^\iota.
\end{equation}
The designation is again borrowed from 3D optics and calls for the
optical path length. For simplicity we will consider only those
cases where
\begin{equation}
    g_0 = \sigma_0, ~~~~ g_i = {n^j}_i \sigma_j;
\end{equation}
to get {$\mathrm{d}s = \sigma_0 \mathrm{d}t + g_i \mathrm{d}x^i$}.
That is, we are considering only spaces where the refractive index
is a 4-rank tensor and does not modify the zeroth frame vector. A
further simplification results from imposing that the refractive
index depends only on the 3 spatial $x^m$ coordinates\footnote{This
restriction is not applied in the appendix.}. We now replace the
null displacement condition by a similar condition applied to
optical displacement. From $\mathrm{d}s^2 = 0$ we write immediately
\begin{equation}
    \mathrm{d} t^2 = g_{ij} \mathrm{d}{x}^i \mathrm{d}{x}^j;
\end{equation}
where $g_{ij} = g_i \dprod g_j$. This is interpreted as the metric
of 4D space with Euclidean signature, or 4DO space.

Multiplying the optical displacement \eqref{eq:opticaldisp} on the
right and on the left by $g^4$, simultaneously replacing $x^4$ by
$\tau$ we obtain
\begin{align}
    \mathrm{d}s g^4 &= \sigma_0 g^4 \mathrm{d}t + g_m g^4
    \mathrm{d}x^m + g_4 g^4  \mathrm{d}\tau;\\
    g^4 \mathrm{d}s &=
    g^4 \sigma_0 \mathrm{d}t + g^4 g_m
    \mathrm{d}x^m + g^4 g_4  \mathrm{d}\tau.
\end{align}
Multiplying the two equations member by member, the first member
becomes $g^{44}\mathrm{d}s^2$ and must therefore be null. We have
then
\begin{equation}
    0 = g^{44} \left(-\mathrm{d}t^2 + g_{mn}\mathrm{d}x^m \mathrm{d}x^n
    + g_{44}\mathrm{d}\tau^2 \right).
\end{equation}
Note that non-scalar terms in the second member cancel out
necessarily, so that the first member can be null. Rearranging the
equation we can write
\begin{equation}
    \mathrm{d}\tau^2 = \frac{1}{g_{44}} \left(\mathrm{d}t^2 -
    {g_{mn}}\mathrm{d}x^m    \mathrm{d}x^n \right).
\end{equation}
This is clearly a GR metric if the second member depends only on the
$x^m$ coordinates, that is, if the metric is static. We have thus
established a metric conversion method between GR and 4DO applicable
to static metrics. The importance of this conclusion cannot be
overstressed; we have concluded that geodesics of 4DO and GR spaces
can be mapped to each other when the refractive index is a function
only of the 3 spatial coordinates. When this happens any dynamics
which can be studied as free fall in GR can also be studied as free
fall in 4DO, providing a different angle of approach to the same
problem.

The geodesics of 4DO space can be found, as in any other space, by
consideration of the Lagrangian \cite{Martin88}
\begin{equation}
\label{eq:lagrangian}
    L = \frac{g_{i j} \dot{x}^i \dot{x}^j}{2} = \frac{1}{2}\ ;
\end{equation}
from the Lagrangian one derives immediately the conjugate momenta
\begin{equation}
    v_i = \frac{\partial L}{\partial \dot{x}^i} = g_{i j} \dot{x}^j.
\end{equation}
Note the use of the lower index ($v_i$) to represent conjugate
momenta while velocity components have an upper index ($v^i$). The
conjugate momenta are the components of the conjugate momentum
vector $v = g^i v_i$, which can be written in two alternative forms
\begin{equation}
    v = g^i v_i = g^i g_{i j} \dot{x}^j = g_j v^j.
\end{equation}
We conclude that conjugate momentum and velocity are the same vector
but their components are referred to the reciprocal and refractive
index frames, respectively. The geodesic equations can now be
written in the form of Euler-Lagrange equations
\begin{equation}
    \dot{v}_i = \partial_i L;
\end{equation}

Space is isotropic if the refractive index does not depend on
direction, and so the three $g_m$ vectors must be related to
$\sigma_m$ by a common scale factor, which may be a function of
position. The scale coefficient for $g_4$ does not need to be the
same as for the other frame vectors and hence we will characterize
an isotropic space by the refractive index frame
\begin{equation}
    g_m = n_r \sigma_m,~~~~ g_4 = n_4 \sigma_4.
\end{equation}
In problems with spherical symmetry we use spherical coordinates and
it must be
\begin{align}
    g_r &= n_r \sigma_r, \\ g_\theta &= n_r r \sigma_\theta, \\
    g_\varphi &= n_r r \sin \theta \sigma_\varphi,  \\ g_4 &= n_4
    \sigma_\tau;
\end{align}
with both $n_r$ and $n_4$ functions of $r$.

We will now look at Schwarzschild's metric to see how it can be
transposed to 4D optics. The usual form of the metric is
\begin{equation}
    \mathrm{d}\tau^2 = \left(1-\frac{2m}{\chi} \right)\mathrm{d}t^2
    -\left(1-\frac{2m}{\chi} \right)^{-1}\mathrm{d}\chi^2 - \chi^2
    \left(\mathrm{d}\theta^2 + \sin^2 \theta \mathrm{d}\varphi^2 \right);
\end{equation}
where $m$ is a spherical mass and $\chi$ is a radial coordinate, not
the distance to the spherical mass' centre. This form is
non-isotropic but a change of coordinates can be made that returns
an isotropic form, see \citet[section 14.7]{Inverno96}.

\begin{equation}
    r=\left(\chi-m+\sqrt{\chi^2-2m \chi}\right)/2;
\end{equation}
and the new form of the metric is
\begin{equation}
    \mathrm{d}\tau^2 =
    \left(\frac{\displaystyle 1-\frac{m}{2r}}{\displaystyle 1+\frac{m}{2r}}\right)^2
    \mathrm{d}t^2 -
    \left(1+ \frac{m}{2r}\right)^4\left[ \mathrm{d}r^2 - r^2 \left(\mathrm{d}\theta^2
    + \sin^2 \theta \mathrm{d}\varphi^2 \right) \right].
\end{equation}
This corresponds to the refractive index coefficients
\begin{equation}
    n_4 = \frac{\displaystyle 1+ \frac{m}{2r}}{\displaystyle 1-\frac{m}{2r}},
    ~~~~ n_r = \frac{\left(\displaystyle 1 + \frac{m}{2r}\right)^3}
    {\displaystyle 1-\frac{m}{2r}},
\end{equation}
which can then be used in 4DO Euclidean space.

We analyse now the constraints on the refractive index so that
experimental data on light bending and perihelium advance in closed
orbits can be predicted; this will allow us to propose another set
of refractive indices which will be more convenient than those just
obtained. Light rays are characterized by $\mathrm{d}\tau=0$ both in
4DO and in general relativity; the effective refractive index for
light is then
\begin{equation}
    \sqrt{\frac{1}{\sum (\dot{x}^m)^2}} = n_r.
\end{equation}
For compatibility with experimental observations $n_r$ must be
expanded in series as (see \cite{Will01})
\begin{equation}
    n_r = 1 +\frac{2 m}{r} + \mathrm{O}(1/r)^2.
\end{equation}
This is the bending predicted by Schwarzschild's metric and has been
confirmed by observations.

For the analysis of orbits its best to rewrite \eqref{eq:lagrangian}
for spherical coordinates. Since we know that orbits are flat we can
make $\theta = \pi/2$
\begin{equation}
\label{eq:orbit}
    n_4^2 \dot{\tau}^2 + n_r^2 (\dot{r}^2 + r^2 \dot{\varphi}^2 ) =1.
\end{equation}
The metric depends only on $r$ and we get two conservation equations
\begin{equation}
    n_4^2 \dot{\tau} = \frac{1}{\gamma},~~~~ n_r^2 r^2 \dot{\varphi}
    = J_\varphi.
\end{equation}
Replacing in \eqref{eq:orbit}
\begin{equation}
    \frac{1}{\gamma^2 n_4^2}\, + n_r^2 \dot{r}^2 +
    \frac{J_\varphi^2}{n_r^2 r^2}\, = 1.
\end{equation}
The solution of this equation calls for a change of variable $r =
1/b$; as a result it is also $\dot{r} =\dot{\varphi}
\mathrm{d}r/\mathrm{d}\varphi$; replacing in the equation and
rearranging
\begin{equation}
    \left(\frac{\mathrm{d}b}{\mathrm{d}\varphi} \right)^2 =
    \frac{n_r^2}{J_\varphi^2}\, - \frac{n_r^2}{J_\varphi^2 \gamma^2 n_4^2}\, -b^2.
\end{equation}
To account for light bending we have established that $n_r \approx
1+2m b$. For $n_4$ we need 2nd order approximation \cite{Will01}, so
we make $n_4 \approx 1 + \alpha m b + \beta m^2 b^2$. We can also
assume that velocities are low, so $\gamma \approx 1$
\begin{equation}
    \left(\frac{\mathrm{d}b}{\mathrm{d}\varphi} \right)^2 \approx
     \frac{2
    \alpha m}{J_\varphi^2}\, b + \left(-1 + \frac{8 \alpha m^2}{J_\varphi^2} -
    \frac{3 \alpha^2 m^2}{J_\varphi^2} + \frac{2 \beta m^2}{J_\varphi^2} \right)
    b^2.
\end{equation}
For compatibility with Kepler's 1st order predictions $\alpha =1$;
then, for compatibility with observed planet orbits, $\beta = 1/2$.
Together with the constraint for $n_0$, these are the conditions
that must be verified by the refractive indices to be in agreement
with experimental data; any refractive indices verifying such
conditions are then perfectly legitimate in terms of predictions for
those two observations.

We know, of course, that the refractive indices corresponding to
Schwarzschild's metric verify the constraints above, however that is
not the only possibility. Schwarzschild's metric is a consequence of
Einstein's equations when one postulates that vacuum is empty of
mass and energy, but the same does not necessarily apply in 4DO. In
\cite{Almeida04:4} we proposed a counterpart to Einstein equations
in 4DO whose solutions are in full agreement with observations; the
resulting refractive index is
\begin{align}
    \label{eq:nr}
    n_r &= \mathrm{e}^{2 m/r} \approx 1 + \frac{2 m}{r}\,;\\
    \label{eq:n0}
    n_4 &= \mathrm{e}^{m /r} \approx 1 + \frac{m}{r}\, +
    \frac{m^2}{2r^2}\,.
\end{align}
\citet{Montanus01} arrives at the same solutions with a different
reasoning; the same metric is also due to Yilmaz
\cite{Yilmaz58,Yilmaz71,Ibison05,Yilmaz05}.

These refractive index coefficients are as effective as those
derived from Schwarzschild's metric for light bending and perihelium
advance prediction for small $m/r$; there is one singularity for
$r=0$ which is not a physical difficulty since before that stage
quantum phenomena have to be considered and the metric ceases to be
applicable; in other words, we must change from geometric to wave
optics approach.
\section{Hyperspherical coordinates}
Deriving physical equations and predictions from purely geometrical
equations is an exercise whose success depends on the correct
assignment of coordinates to physical entities; the same space will
produce different predictions if different options are taken for
coordinate assignment. Since the birth of special relativity it has
been usual to assign three coordinates to orthogonal directions in
physical space and a zeroth coordinate to time. This is a totally
arbitrary assignment, which has gained acceptance by the correct
predictions it originates in many circumstances. We discussed above
that it is also perfectly legitimate to replace the assignment of
coordinate zero to time by an assignment of coordinate four with
proper time. Geodesic (straight line) movement can be predicted
equally well in both cases. In terms of curvature, flat space is
usually associated with absolute emptiness in a physical sense.

We are now going to experiment with a different assignment of flat
space coordinates, which will explore the possibility that physics
and the Universe have an inbuilt hyperspherical symmetry. The
exercise consists on assigning coordinate $x^4 = \tau$ to the radius
of an hypersphere and the three $x^m$ coordinates to distances
measured on the hypersphere surface. If the hypersphere radius is
very large we will not be able to notice the curvature on everyday
phenomena, in the same way as everyday displacements on Earth don't
seem curved to us; but the Universe as a whole will manifest the
consequences of its hyperspherical symmetry. Using the Earth as a
3-dimensional analogue of an hyperspherical Universe, although our
everyday life is greatly unaffected by Earth's curvature the
atmosphere senses this curvature and shows manifestations of it in
winds and climate. What we propose here is an exercise; it is an
arbitrary assignment between coordinates and physical entities; the
validity of such exercise can only be judged by the predictions it
allows and how well they conform with observations.

Hyperspherical coordinates are characterized by one distance
coordinate, $\tau$ and three angles $\rho, \theta, \varphi$;
following the usual procedure we will associate with these
coordinates the frame vectors $\{\sigma_\tau, \sigma_\rho,
\sigma_\theta, \sigma_\varphi\}$. The position vector for one point
in 5D space is quite simply
\begin{equation}
    x = t \sigma_0 + \tau \sigma_4.
\end{equation}
In order to write an elementary displacement $\mathrm{d}x$ we must
consider the rotation of frame vectors, but we don't need to think
hard about it because we can extend what is known from ordinary
spherical coordinates \cite{Peterson01}.
\begin{equation}
    \label{eq:hyperdisplacement}
    \mathrm{d}x = \sigma_0 \mathrm{d}t + \sigma_4 \mathrm{d}\tau + \tau \sigma_\rho
    \mathrm{d}\rho + \tau \sin \rho \sigma_\theta \mathrm{d}\theta +
    \tau \sin \rho \sin \theta \sigma_\varphi \mathrm{d}\varphi.
\end{equation}
Just as before, we consider only null displacements to obtain time
intervals;
\begin{equation}
    \mathrm{d}t^2 = \mathrm{d}\tau^2 + \tau^2 \left[\mathrm{d}\rho^2 +
    \sin^2 \rho \left(\mathrm{d}\theta^2 +
    \sin^2 \theta \mathrm{d}\varphi^2 \right)\right].
\end{equation}
The velocity vector, $v = \dot{x} - \sigma_0$, can be immediately
obtained from the displacement vector dividing by $\mathrm{d}t$
\begin{equation}
    \label{eq:velhspher}
    v = \sigma_0 \dot{\tau} + \tau \sigma_\rho
    \dot{\rho} + \tau \sin \rho \sigma_\theta \dot{\theta} +
    \tau \sin \rho \sin \theta \sigma_\varphi \dot{\varphi}.
\end{equation}

Geodesics of flat space are naturally straight lines, no matter
which coordinate system we use, however it is useful to derive
geodesic equations from a Lagrangian of the form
\eqref{eq:lagrangian}; in hyperspherical coordinates the Lagrangian
becomes
\begin{equation}
   2 L = v^2 = \dot{\tau}^2 + \tau^2 \left[\dot{\rho}^2 + \sin^2
   \rho \left(\dot{\theta}^2 + \sin^2 \theta \dot{\varphi}^2 \right)
   \right].
\end{equation}
Because de Lagrangian is independent of $\varphi$ we can establish a
conserved quantity
\begin{equation}
    J_\varphi = \tau^2 \sin^2 \rho \sin^2 \theta \dot{\varphi}.
\end{equation}

It may seem strange that any physically meaningful relation can be
derived from the simple coordinate assignment that we have made,
that is, proper time is associated with hypersphere radius and the
three usual space coordinates are assigned to distances on the
hypersphere radius. This unexpected fact results from the
possibility offered by hyperspherical coordinates to explore a
symmetry in the Universe that becomes hidden when we use Cartesian
coordinates. In the real world we measure distances between objects,
namely cosmological objects, rather than angles; we have therefore
to define a distance coordinate, which is obviously $r = \tau \rho$.
It does not matter where in the Universe we place the origin for $r$
and we find it convenient to place ourselves on the origin.

Radial velocities $\dot{r}$ measure movement in a radial direction
from our observation point; we are particularly interested in this
type of movement in order to find a link to the Hubble relation.
Applying the chain rule and then replacing $\rho$
\begin{equation}
    \dot{r} = \rho \dot{\tau} + \dot{\rho} \tau =
    \frac{\dot{\tau}}{\tau}\, r + \dot{\rho} \tau.
\end{equation}
We expect objects that have not suffered any interaction to move
along $\sigma_\tau$; from \eqref{eq:velhspher} we see that this
implies $\dot{\rho} = \dot{\theta} = \dot{\varphi}=0$ and then
$\dot{\tau}$ becomes unity. Replacing in the equation above and
rearranging
\begin{equation}
    \label{eq:firsthubble}
    \frac{\dot{r}}{r} = \frac{1}{\tau}.
\end{equation}
What this equation tells us is exactly what is expressed by the
Hubble relation. The value of $\tau$ can be taken as constant for
any given observation because the distance information is carried by
photons and these preserve proper time\footnote{In order to preserve
proper time photons must travel on the hypersphere surface and thus
don't follow geodesics; the way in which this is made compatible
with electromagnetism is briefly discussed in the appendix.}. The
first member of the equation is the definition of the Hubble
parameter and we can then write $H = 1/\tau$. In this way we find
the physical meaning of coordinate $\tau$ as being the Universe's
age.

How does the use of hyperspherical coordinates affect dynamics in
our laboratory experiments? We would like to know if these
coordinates need only be considered in problems of cosmological
scale or, on the contrary, there are implications for everyday
experiments. The answer implies rewriting
\eqref{eq:hyperdisplacement} with distance rather than angle
coordinates; replacing $\rho$,
\begin{equation}
    \label{eq:dxhyper}
    \mathrm{d}x = \sigma_0 \mathrm{d}t + \left(\sigma_4 - \frac{r}{\tau}\,
    \sigma_\rho
    \right) \mathrm{d}\tau + \sigma_\rho \mathrm{d}r + r (\sigma_\theta
    \mathrm{d}\theta + \sin \theta \sigma_\varphi
    \mathrm{d}\varphi ).
\end{equation}
Evaluating time intervals from the null displacement condition, as
before
\begin{equation}
    \mathrm{d}t^2 = \left[1+ \left(\frac{r}{\tau}\right)^2 \right]\mathrm{d}\tau^2
    - 2 \frac{r}{\tau}\, \mathrm{d}\tau
    \mathrm{d}r + \mathrm{d}r^2 + r^2 (\mathrm{d}\theta^2 + \sin^2
    \theta \mathrm{d}\varphi^2).
\end{equation}
This would be a version of \eqref{eq:twospaces1} in spherical
coordinates, were it not for the extra terms with powers of $r/\tau$
in the second member. The coefficient $r/\tau$ implies a comparison
between the distance from the object to the observer and the size of
the Universe; remember that $\tau$ is both time and distance in
non-dimensional units. We can say that ordinary special relativity
will apply for objects which are near us, but distant objects will
show in their movement an effect of the Universe's hyperspherical
nature.

We have established the refractive indices $n_r$ \eqref{eq:nr} and
$n_4$ \eqref{eq:n0} to account for the dynamics near a massive
sphere using Cartesian coordinates; since this is frequently applied
on a cosmological scale, we must find out how the dynamics is
modified by the use of hyperspherical coordinates. Using the
refractive indices and hyperspherical coordinates, noting that $n_r
= n_4^2$, the optical displacement \eqref{eq:opticaldisp} becomes
\begin{equation}
    \mathrm{d}s = \sigma_0 \mathrm{d}t + n_4 \sigma_4 \mathrm{d}
    \tau + n_4^2 \left(\tau \sigma_\rho \mathrm{d} \rho + \tau \sin
    \rho \sigma_\theta \mathrm{d} \theta + \tau \sin \rho \sin
    \theta \sigma_\varphi \mathrm{d} \varphi \right).
\end{equation}
In radial displacements we can set $\dot{\theta} = \dot{\varphi} =
0$; introducing this and dividing by $\mathrm{d}t$
\begin{equation}
    \label{eq:dots}
    \dot{s} = \sigma_0 + n_4 \sigma_4 \dot{\tau} + n_4^2 \tau \sigma_\rho
    \dot{\rho}.
\end{equation}
Squaring $\dot{s}$ and invoking null displacement condition
\begin{equation}
    \label{eq:radvel}
    n_4^2 \dot{\tau}^2 + n_4^4 \tau^2 \dot{\rho}^2 = 1.
\end{equation}
and replacing $\tau \dot{\rho}$ by $\dot{r}- r \dot{\tau}/\tau$
\begin{equation}
    n_4^2 \dot{\tau}^2 + n_4^4 \left[\dot{r}^2 + \left(
    \frac{\dot{\tau}}{\tau} \right)^2 r^2 - 2 \dot{\tau} \dot{r}
    \frac{r}{\tau} \right] = 1.
\end{equation}
Dividing both members by $n_4^4 r^2$ and rearranging results in the
equation
\begin{equation}
    \label{eq:dotrr}
    \left(\frac{\dot{r}}{r} \right)^2 =  \left(
    \frac{1}{n_4^4} - \frac{\dot{\tau}^2}{n_4^2}
    \right)\frac{1}{r^2} -\left(\frac{\dot{\tau}}{\tau} \right)^2
     + 2 \frac{\dot{\tau} \dot{r}}{\tau r}.
\end{equation}
As a further step we expand the second member in series of $m$ and
take the two first terms, in order to get an equation that allows
comparison to those used in cosmology.
\begin{equation}
    \label{eq:newfriedman}
    \left(\frac{\dot{r}}{r} \right)^2 \approx \frac{1 - \dot{\tau}^2}{r^2}
    + \frac{(2 \dot{\tau}^2
    -4) m}{r^3} -\left(\frac{\dot{\tau}}{\tau}
    \right)^2 + 2 \frac{\dot{\tau} \dot{r}}{\tau r}.
\end{equation}
The previous equation applies to bodies moving radially under the
influence of mass $m$ located at the origin which is, remember, the
observer's position. For comparison we derive the corresponding
equation in Cartesian coordinates; starting with \eqref{eq:radvel}
it is now
\begin{equation}
    n_4^2 \dot{\tau}^2 + n_4^4 \dot{r}^2 =1;
\end{equation}
dividing by $n_4^4 r^2$ and rearranging
\begin{equation}
    \left(\frac{\dot{r}}{r} \right)^2 = \left(
    \frac{1}{n_4^4} - \frac{\dot{\tau}^2}{n_4^2}
    \right)\frac{1}{r^2} \approx \frac{1 - \dot{\tau}^2}{r^2} + \frac{(2 \dot{\tau}^2
    -4) m}{r^3}.
\end{equation}

If we want to apply these equations to cosmology it is easiest to
follow the approach of Newtonian cosmology, which produces basically
the same results as the relativistic approach but presumes that the
observer is at the centre of the Universe
\cite{Inverno96,Narlikar02}. In order to adopt a relativistic
approach we need equations that replace Einstein's in 4DO. A set of
such equations has been proposed \cite{Almeida04:4} but their
application in cosmology has not yet been tested, so we will have to
defer this more correct approach to a forthcoming paper. The
strategy is to consider a general object at distance $r$ from the
observer, moving away from the latter under the gravitational
influence of the mass included in a sphere of radius $r$. If we
designate by $\mu$ the average mass density in the Universe, then
mass $m$ in \eqref{eq:newfriedman} is $4 \pi \mu r^3/3$; this will
have to be considered further down.

Friedman equation governs standard cosmology and can be derived both
from Newtonian and relativistic dynamics, with different
consequences in terms of the overall size of the Universe and the
observer's privileged position. From the cited references we write
Friedman equation as
\begin{equation}
    \left(\frac{\dot{r}}{r} \right)^2 = \frac{8 \pi}{3}\, \mu +
    \frac{\Lambda}{3 } - \frac{k }{r^2};
\end{equation}
with $\Lambda$ a cosmological constant and $k$ the curvature
constant; the gravitational constant was not included because it is
unity in non-dimensional units and the equation is written in real,
not comoving, coordinates. In order to compare
\eqref{eq:newfriedman} with Friedman equation there is a problem
with the last term because the Hubble parameter $\dot{r}/r$ does not
appear isolated in the first member; we will find a way to
circumvent the problem later on but first let us look at what
\eqref{eq:newfriedman} tells us when the mass density is zeroed. In
this case $n_4 = 1$ and we find from \eqref{eq:radvel} that
$\dot{\tau}$ is unity, unless $\dot{\rho}$ is non-zero, for which we
can find no reasonable explanation. Replacing $n_4$ and $\dot{\tau}$
with unity in \eqref{eq:newfriedman} we find that $\dot{r}/r =
1/\tau$, confirming what had already been found in
\eqref{eq:firsthubble}. Comparing with Friedman equation, this
corresponds to a flat Universe with a critical mass density $\mu =
\mu_c$; it is immediately obvious that $\mu_c = 3 /(8 \pi \tau^2)$.
Let us not overlook the importance of this conclusion because it
completely removes the need for a critical density if the Universe
is flat; remember this is one of the main reasons to invoke dark
matter in standard cosmology. Notice also that this conclusion does
not depend on a privileged observer, because it is just a
consequence of space symmetry and not of dynamics.

Let us now see what happens when we consider a small mass density;
here we are talking about matter that is observed or measured in
some way but not postulated matter. The matter density that we will
consider is of the order of 1\% of the presently accepted value. It
is therefore just a perturbation of the flat solution that we
described above and the fact that we are presuming a privileged
observer has to be taken just for this perturbation. The first thing
we note when we consider matter density is that $\dot{\tau} <1$,
because there is now a component of the velocity vector along
$\sigma_\rho$. Ideally we should solve the Euler-Lagrange equations
resulting from \eqref{eq:radvel} in order to find $\dot{\tau}$ and
$\dot{\rho}$ but this is a difficult process and we shall carry on
with just a qualitative discussion. Considering that we are
discussing a perturbation it is legitimate to make $\dot{r}/{r}
\approx \dot{\tau}/\tau$ and the two last terms in the second member
of \eqref{eq:newfriedman} can be combined into one single term
$(\dot{\tau}/\tau)^2$, the same as we encountered for the flat
solution, albeit with a numerator slightly smaller than unity. The
first term has now become slightly positive and we can see from
Friedman equation that this corresponds to a negative curvature
constant, $k$, and to an open Universe. Lastly the second term
includes the mass $m$ of a sphere with radius $r$ and can be
simplified to $8 \pi \mu (\dot{\tau}^2 - 2)/3$; this has the effect
of a negative cosmological constant; the combined effect of the two
terms is expected to close the Universe \cite{Martin88,Narlikar02}.
The previous discussion was done in qualitative terms, making use of
several approximations, for which reason we must question some of
the findings and expect that after more detailed examination they
may not be quite as anticipated; in particular there is concern
about the refractive indices used, which were derived in Cartesian
coordinates both by the author and those that preceded him in using
an exponential metric; it may happen that the transposition to
hyperspherical coordinates has not been properly made, with
consequences in the perturbative analysis that was superimposed on
the flat solution. The latter, however, is totally independent of
such concerns and allows us to state that the assumption of
hyperspherical symmetry for the Universe dispenses with dark matter
in accounting for the gross of observed expansion.

Dark matter is also called in cosmology to account for the extremely
high rotation velocities found in spiral galaxies
\cite{Silk97,Rubin78} and we will now take a brief look at how
hyperspherical symmetry can help explain this phenomenon. Galaxy
dynamics is an extremely complex subject, which we do not intend to
explore here due to lack of space but most of all due to lack of
author's competence to approach it with any rigour; we will just
have a very brief outlook at the equation for flat orbits, to notice
that an effect similar to the familiar Coriollis effect on Earth can
arise in an expanding hyperspherical Universe and this could explain
most of the observed velocities on the periphery of galaxies. Let us
recall \eqref{eq:dxhyper}, divide by $\mathrm{d}t$ and invoke null
displacement to obtain the velocity
\begin{equation}
    v = \left(\sigma_4 -
    \frac{r}{\tau}\,
    \sigma_\rho
    \right) \dot{\tau} + \sigma_\rho \dot{r} + r (\sigma_\theta
    \dot{\theta} + \sin \theta \sigma_\varphi
    \dot{\varphi} ).
\end{equation}
If orbits are flat we can make $\theta = \pi/2$ and the equation
simplifies to
\begin{equation}
    \label{eq:rotvel}
    v = \dot{\tau} \sigma_4
    +  \left( \dot{r} -
    \frac{r \dot{\tau}}{\tau} \right) \sigma_\rho + r
    \dot{\varphi} \sigma_\varphi.
\end{equation}
Suppose now that something in the galaxy is pushing outwards
slightly, so that the parenthesis is zero; this happens if
$\dot{r}/r = \dot{\tau}/\tau$ and can be caused by a pressure
gradient, for instance. The result is that \eqref{eq:rotvel} now
accepts solutions with constant $r \dot{\varphi}$, which is exactly
what is observed in many cases; swirls will be maintained by a
radial expansion rate which exactly matches the quotient
$\dot{\tau}/\tau$. In any practical situation $\dot{\tau}$ will be
very near unity and the quotient will be virtually equal to the
Hubble parameter; thus the expansion rate for sustained rotation is
$\dot{r}/r \approx H$. If applied to our neighbour galaxy Andromeda,
with a radial extent of $30~\mathrm{kpc}$, using the Hubble
parameter value of $81~\mathrm{km}~\mathrm{ s}^{-1}/\mathrm{Mpc}$,
the expansion velocity is about $2.43~\mathrm{km}~\mathrm{ s}^{-1}$;
this is to be compared with the orbital velocity of near
$300~\mathrm{km}~ \mathrm{s}^{-1}$ and probably within the error
margins. An expansion of this sort could be present in many galaxies
and go undetected because it needs only be of the order of 1\% the
orbital velocity.

\section{Conclusion}

The approach to the equations that govern the Universe examined in
this paper can be compared to the revolution brought to 15th century
geography and navigation by the consideration of a spherical Earth,
a concept as old as Pythagoras and Aristotle but not widely accepted
until then. It is unimaginable today to explain any world scale
phenomenon without recourse to spherical coordinates, because these
make full exploitation of Earth's spherical symmetry and render
equations enormously simpler than they would be if expressed in
Cartesian coordinates.

Making an hypothesis that the Universe as a whole has an inbuilt
hyperspherical symmetry we were able to derive Hubble's law as a
direct consequence of geometry in a Universe completely devoid of
any matter. The existence of a minute mass density can then be seen
to introduce a perturbation in the main picture, being responsible
for a slight curvature and a small cosmological constant. Similarly
to what happens in the Earth's atmosphere, we were also able to
demonstrate the existence of constant rotation velocity swirls that
can be the basis for understanding galaxy dynamics.

Mathematically the argument was set on purely geometrical grounds,
with 5-dimensional spacetime as a point of departure. This space was
shown to produce both GR spacetime and an Euclidean metric
4-dimensional space, by an imposition of null displacement.
Euclidean metric 4D space was then used to formulate the
hyperspherical symmetry hypothesis and derive its consequences.

\bibliographystyle{CCC}
\bibliography{Abrev,aberrations,assistentes}

\appendix

\section{Electromagnetism in 5D spacetime}

We will treat electromagnetism as a local phenomenon, avoiding the
need to use hyperspherical coordinates. The easiest way to include
electromagnetism in the geometry of 5-dimensional spacetime is to
consider a non-orthonormed frame; in terms of the reciprocal frame
we make
\begin{equation}
    g^\mu = \sigma_\mu,~~~~ g^4 = \frac{q}{m}\, A_\mu \sigma^\mu + \sigma^4.
\end{equation}
This is used for the definition of a covariant derivative
\begin{equation}
    \mathcal{D} = g^\iota \partial_\iota = \sigma^\mu \partial_\mu +
    (\sigma_4 + \frac{q}{m}\, A_\mu \sigma^\mu) \partial_4.
\end{equation}
A covariant Laplacian is the defined as $\mathcal{D}^2 = \mathcal{D}
\dprod \mathcal{D}$ and being the square of a vector it is a scalar.
It follows that the Laplacian of a vector must always be a vector
and we have by necessity
\begin{equation}
    \mathcal{D}^2 g^4 = \frac{J}{m} .
\end{equation}

The covariant derivative of $g^4$ can have scalar and bivector parts
but by choice of $A_\mu$ we can zero the scalar part and thus define
the Faraday bivector
\begin{equation}
    F = m \mathcal{D} \wprod g^4,
\end{equation}
so that $\mathcal{D} F = J$, our version of Maxwell's equations
\cite{Doran03}. In the absence of currents we look for solutions
with the second member zero, that is $\mathcal{D}^2 g^4 = 0$, which
reduces to
\begin{equation}
    \mathcal{D}^2 A = 0,
\end{equation}
with $A = A_\mu \sigma^\mu = A^\mu \sigma_\mu$. If $A$ does not
depend on $x^4$, then the Laplacian reduces to $-
\partial^2/\partial t^2 + \sum \partial^2/\partial (x^m)^2$ and the
equation is a straightforward wave equation with plane wave
solutions normal to $\sigma_4$.

We use the argument above to sustain that electromagnetic waves must
follow lines normal to $\sigma_4$ in any circumstance, even when
$x^4$ is the radius of an hypersphere. Electromagnetic waves will
follow geodesics in usual flat space but they will follow great
circles on the hypersphere if hyperspherical symmetry is assumed.

\end{document}